\documentclass[a4paper,11pt]{article}
\usepackage{pos}
\usepackage{cleveref}
\usepackage{siunitx}
\usepackage{adjustbox}
\usepackage{csquotes}
\usepackage{overpic}
\usepackage{isotope}
\usepackage[symbol]{footmisc}
\usepackage[numbers]{natbib}
\DeclareSIUnit[quantity-product = {}]
\speedoflight{\text{c}}
\usepackage{wrapfig}
\usepackage{caption}
\usepackage{subcaption}

\title{POLAR-2 – Latest Developments of the Next Generation GRB Polarimeter}

\author[a]{Johannes Hulsman}
\author*[a]{Philipp Azzarello}
\author[b]{Jörg Bayer}
\author[a]{Franck Cadoux}
\author[b]{Mariachiara Celato}
\author[c]{Nicolas De Angelis}
\author[a]{Yannick Favre}
\author[b]{Aaron Feder}
\author[d]{Jochen Greiner}
\author[b]{Alejandro Guzman}
\author[a]{Coralie Husi}
\author[a]{Vishal Kumar}
\author[e]{Hancheng Li}
\author[a]{Mobin Mobaseri}
\author[a]{Gabriel Pelleriti}
\author[f]{Agnieszka Pollo}
\author[e]{Nicolas Produit}
\author[f]{Dominik Rybka}
\author[b]{Andrea Santangelo}
\author[g]{Jianchao Sun}
\author[b]{Chris Tenzer}
\author[a]{Xin Wu}
\author[g,h]{Shuang-Nan Zhang}

\affiliation[a]{DPNC, University of Geneva, Quai Ernest-Ansermet 24, 1205 Geneva, Switzerland}
\affiliation[b]{Institut für Astronomie und Astrophysik Tübingen, Universität Tübingen, Sand 1, D-72076 Tübingen, Germany}
\affiliation[c]{INAF-IAPS, via del Fosso del Cavaliere 100, 00133 Rome, Italy}
\affiliation[d]{Max-Planck Institute for Extraterrestrial Physics, Giessenbachstr. 1, 85748 Garching, Germany}
\affiliation[e]{Geneva Observatory, ISDC, University of Geneva, Chemin d'Ecogia 16, 1290 Versoix, Switzerland}
\affiliation[f]{National Centre for Nuclear Research, ul. A. Sołtana 7, 05-400 Otwock, Świerk, Poland}
\affiliation[g]{Key Laboratory of Particle Astrophysics, Institute of High Energy Physics, Chinese Academy of Sciences, 100049 Beijing, China}
\affiliation[h]{University of Chinese Academy of Sciences, 100049 Beijing, China}

\emailAdd{philipp.azzarello@unige.ch}

\abstract{Gamma-Ray Bursts (GRBs) are among the most energetic events in the Universe. Despite over 50 years of research and measurements their prompt emission remains poorly understood, with key questions surrounding the structure of relativistic jets, magnetic field configurations, and dominant radiation mechanisms. Polarization measurements are critical in resolving these uncertainties. The POLAR mission, operational in 2016-2017 on Tiangong-2, provided the most statistically significant GRB polarization data. Its results indicated low time-averaged polarization with hints of temporal evolution. However, POLAR’s limited sensitivity, small effective area, and restricted energy range prevented more detailed time- and energy-resolved analyses in addition to a larger sample of GRB polarization measurements. POLAR-2 is designed to address these limitations by offering a fourfold increase in effective area (at least) and an extended energy range of 30–800 keV by utilizing Silicon Photomultipliers (SiPMs) and an updated module design, enabling the differentiation of competing GRB emission models. The instrument comprises of 100 polarimeter modules (each with 64 plastic scintillator bars), wherein the polarization angle is extracted through Compton Scattering of the gammas. The polarimeter module design was validated during an ESRF beam test campaign in 2023. The instrument was developed by a joint effort of Switzerland, China, Poland and Germany and is planned for launch in early 2028. Currently, POLAR-2 is in its production phase with the first module targets being produced. We will provide an overview of the current status of the development.}

\ConferenceLogo{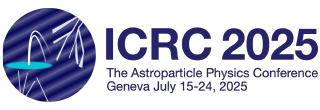}

\FullConference{39th International Cosmic Ray Conference (ICRC2025)\\
 15-24 July 2025\\
Geneva, Switzerland\\}

\begin{document}
\maketitle

\section{Introduction}
\subsection{Gamma-ray Bursts}

Gamma-ray bursts (GRBs) are the universe’s brightest high-energy photon events, releasing up to \qty{\approx e47}{\joule}  over timescales ranging from milliseconds to minutes \cite{galaxies9040082}. Polarization measurements are crucial for understanding these mechanisms \cite{lundman2016polarizationgammarayburstsdissipative}, as the polarization degree (PD) and polarization angle (PA) encode information about the jet's composition, magnetic field configuration, energy dissipation processes and angular structure\cite{galaxies9040082}. Early polarization measurements faced instrumental limitations. However, recent dedicated missions, such as POLAR \cite{PRODUIT2018259} and GAP \cite{Yonetoku2011}, provide more precise polarization measurements, significantly advancing our understanding of these powerful cosmic phenomena.

\subsection{Lessons from POLAR}

The POLAR instrument was launched in September 2016 on the Tiangong-2 and (at its time) measured the most precise GRB polarization in the \qtyrange{50}{500}{\keV} range. It derived the polarization information by exploiting the dependence of the Compton scattering angle to the polarization vector (as described by the Klein-Nishina equation) \cite{Produit2023}. POLAR was composed of 25 polarimeter modules, each containing an \numproduct{8x8} EJ-248M scintillator bars coupled to MaPMT readouts. This segmented design enabled it to extract the PD and PA from the recorded scattering events with greater sensitivity than previous instruments\cite{Produit_2018}. During POLAR's 6-month operation, it detected 55 GRBs jointly with other instruments, 14 of which yielded high-quality polarimetric data \cite{Kole2020}. Furthermore, thanks to its improved sensitivity, POLAR made it possible to study the time dependence of GRB polarization, thereby exposing new aspects of the temporal evolution of GRB emission. Most of POLAR's time-integrated measurements returned low PDs at $\sim$10\%, consistent with the dissipative photosphere scenario \cite{Zhang2019}. However, time-resolved analysis of GRB 170114A revealed PDs of $\sim$30\% with varying PAs, suggesting that the low or moderate time-averaged polarization may arise from integrating polarized emission with evolving PAs over time \cite{Burgess2019, Zhang2019, Kole2020}. These findings underscore the necessity of time-resolved polarimetry for probing jet dynamics. In addition to temporal behavior, POLAR explored energy-dependent polarization trends, although the limited statistics prevented any definitive conclusion \cite{DeAngelis:2023E/}. Based on POLAR's findings, it was understood that there is a need to develop an instrument which will provide a large sample of GRB measurements through a longer mission profile while having the capacity to provide better statistics and sensitivity for time and energy-resolved polarimetry analyses. These design requirements are taken into account for POLAR-2.

\section{The POLAR-2 Mission}

\subsection{Overview}

POLAR-2 is a follow-up mission to the successful POLAR mission, dedicated to deliver exceptional sensitivity and unparalleled precision in measuring the linear polarization of gamma ray bursts (GRBs). It is headed by the University of Geneva, including partners from Poland, Germany and China. The instrument will be installed on the China Space Station (CSS). Based on the current Fermi GRB catalogue, the mission is projected to detect at least 200 GRBs per year \cite{Produit2023}, providing critical insight into the magnetic field topologies and energy dissipation mechanisms that operate within GRB jets. POLAR-2's wide field of view and envisioned rapid alert system also make it ideally suited for identifying GRBs that coincide with gravitational wave (GW) events, thereby enhancing its value in multimessenger astronomy. This functionality aligns with the mission's goal of serving as a cornerstone for collaborative astrophysical investigations. 

\begin{wrapfigure}{r}{0.5\textwidth}
  \begin{center}
  \includegraphics[width=0.48\textwidth]{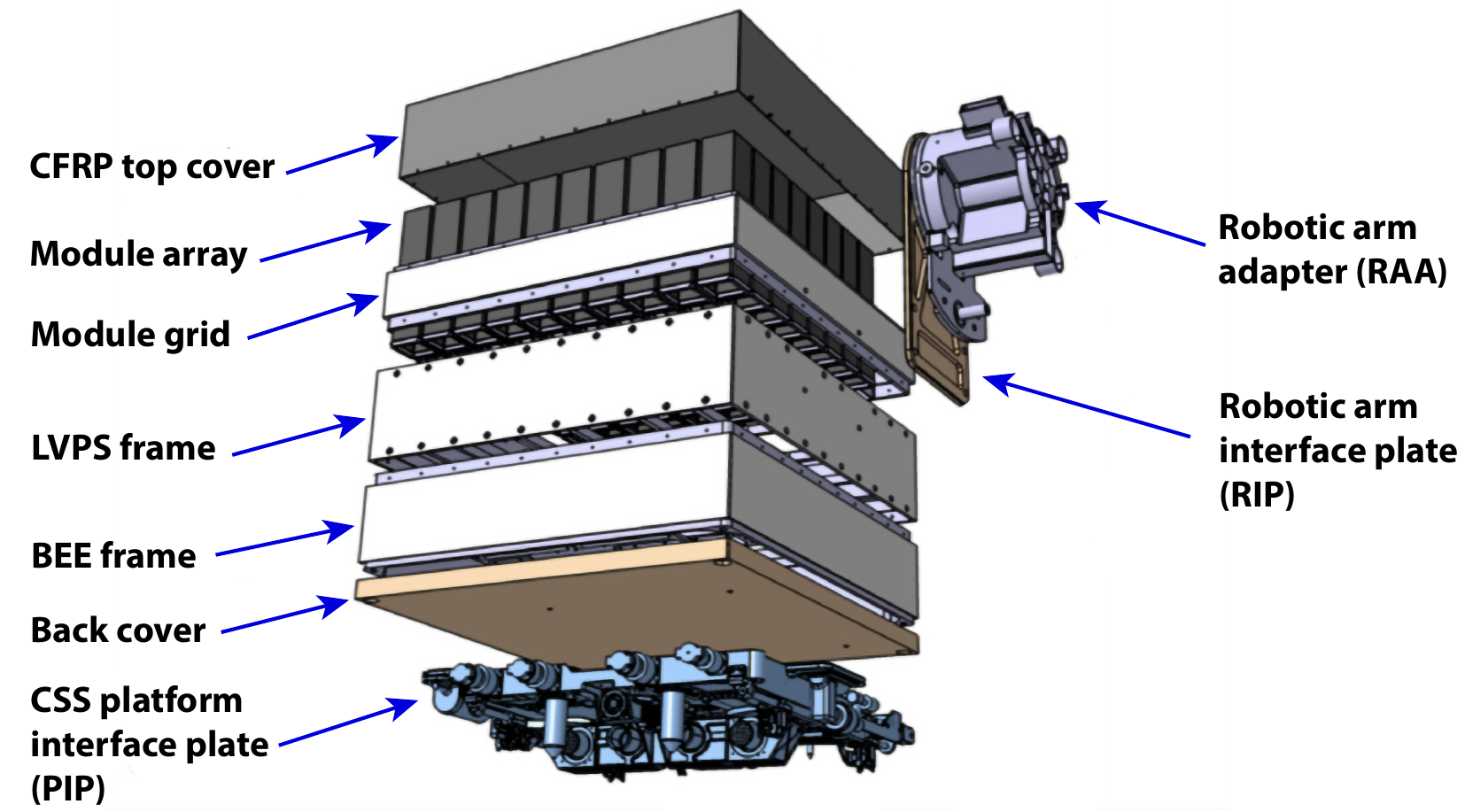}
  \end{center}
  \caption{Exploded view of the POLAR-2 payload}
  \label{fig:POLARinstrument}
\end{wrapfigure}

\subsection{Instrument Design}

\begin{figure}
     \centering
     \begin{subfigure}[b]{\textwidth}
         \centering
         \includegraphics[height=4.5cm]{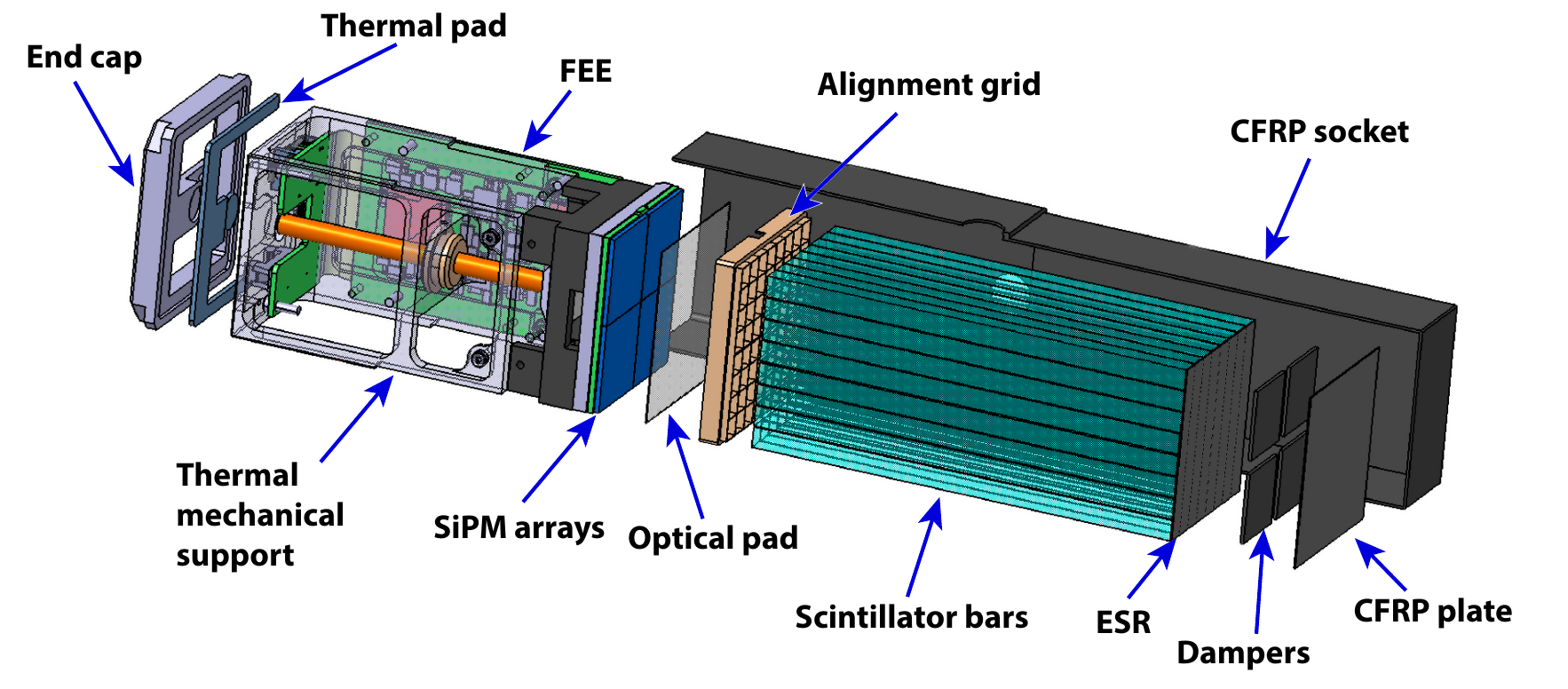}
         \caption{}
         \label{fig:POLAR2ModuleFee}
     \end{subfigure}
     \hfill
     \begin{subfigure}[b]{\textwidth}
         \centering
         \includegraphics[scale=0.1]{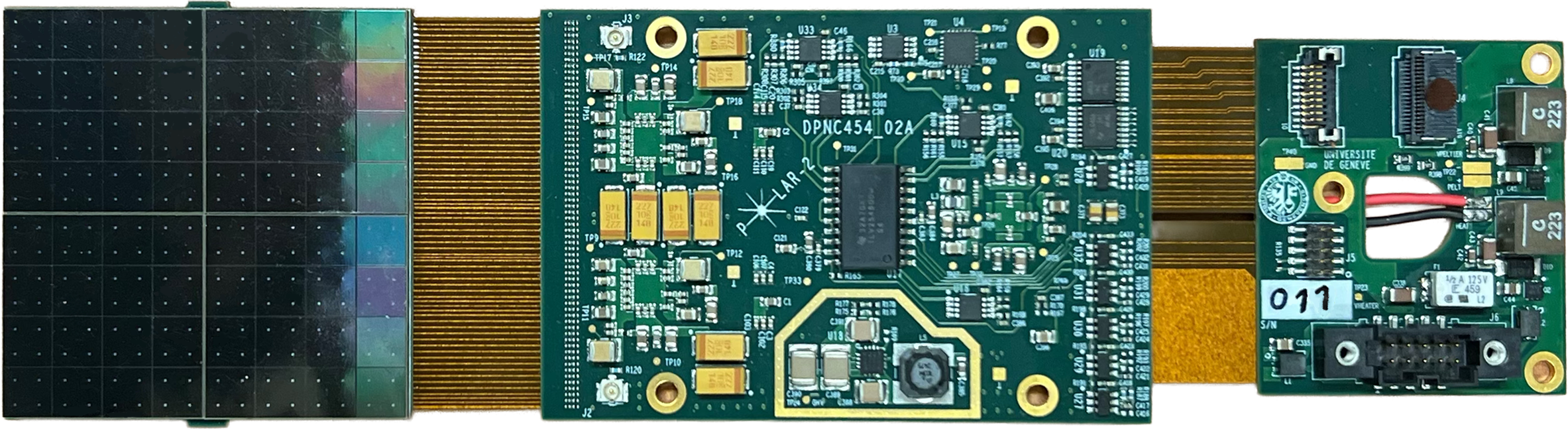}
         \caption{}
         \label{fig:POLAR2FEE}
     \end{subfigure}
    \caption{\textbf{a)} POLAR-2 module exploded view. It consists of 64 scintillator bars, read out by four SiPM arrays connected to a front-end electronics mounted on a thermal mechanical support. Scintillators and FEE are inserted into a CFRP socket (figure taken with permissions from \cite{de_angelis_2023_thesis}). \textbf{b)} Photo of the POLAR-2 front-end electronics (FEE).}
    \label{fig:POLAR2Module}
\end{figure}

POLAR-2 has been designed specifically to overcome the limitations encountered by POLAR (sensitivity, precision and statistics) and technically exceed anything done so far in the field of GRB polarization measurements. It fills the gap by offering up to four times higher overall sensitivity with nearly tenfold improvement below 200 keV and employing state-of-the-art technology to explore GRBs in greater detail with a novel front-end electronic (FEE) for fast event processing and selection \cite{Kole_2025}. \Cref{fig:POLARinstrument} (right) shows an exploded view of the payload. The instrument features a  \numproduct{10x10} array of detector modules held by a grid on the top part. The grid is situated above the low voltage power supplies (LVPS) and the back-end electronics (BEE). Below the BEE is an interface plate to mount the instrument on the CSS (power, data and active cooling) while the robotic arm adapter is needed to move the payload from the cargo hold to its final position. Each polarimeter module contains 64 scintillator bars (EJ-248M) of dimension \qtyproduct[product-units = power]{125x5.9x5.9}{\mm}. The light produced in the scintillators are read out by four 16-channel Hamamatsu SiPMs (S13361-6075NE-04) whose signal is processed by a dedicated FEE (see \Cref{fig:POLAR2FEE}). To maximize the light collection, the bars are wrapped with two layers of reflective foils \cite{DeAngelis_2025}. 

The FEE (\Cref{fig:POLAR2FEE}) is a custom design \cite{Kole_2025} to read out the 64 SiPM channels.  It is composed of three PCB elements, connected with flex cables. The left part hosts the four SiPM arrays, the central part the readout electronics (on the bottom side) and the bias voltage control (top side). The communication and power connectors are mounted on the right part. The readout, whose SiPM signals are read out by two CITIROC ASICs, is controlled by a Microsemi IGLOOv2 FPGA. In normal operation mode, the FEE power consumption is \qty{2.0}{\watt}. The PCB is designed to reduce the electronics heat transfer to the SiPM arrays, keeping the temperature as low and stable as possible. Three sensors monitor the SiPM temperatures to adapt the operating bias voltage, while keeping the overvoltage constant. Finally, the heaters are placed below the SiPMs, to perform periodic annealing and thus limit the effects of radiation damage \cite{DEANGELIS2023167934}. The event selection criteria depends on the charge and arrival timing information in each channel. As the SiPMs are at the core of POLAR-2’s design (upgrading from POLAR's MaPMTs), the detection efficiency has been therefore enhanced, with a photon detection efficiency of 50\% compared to POLAR’s 20\%\cite{Produit2023}. These technological leaps (SiPM technology and a faster readout architecture) enable POLAR-2 to detect gamma rays across a wider energy range, with significantly improved sensitivity below 200 keV. 

\section{Expected Scientific Performance}

\begin{figure}
     \centering
     \begin{subfigure}[b]{0.48\textwidth}
         \centering
         \includegraphics[scale=0.3]{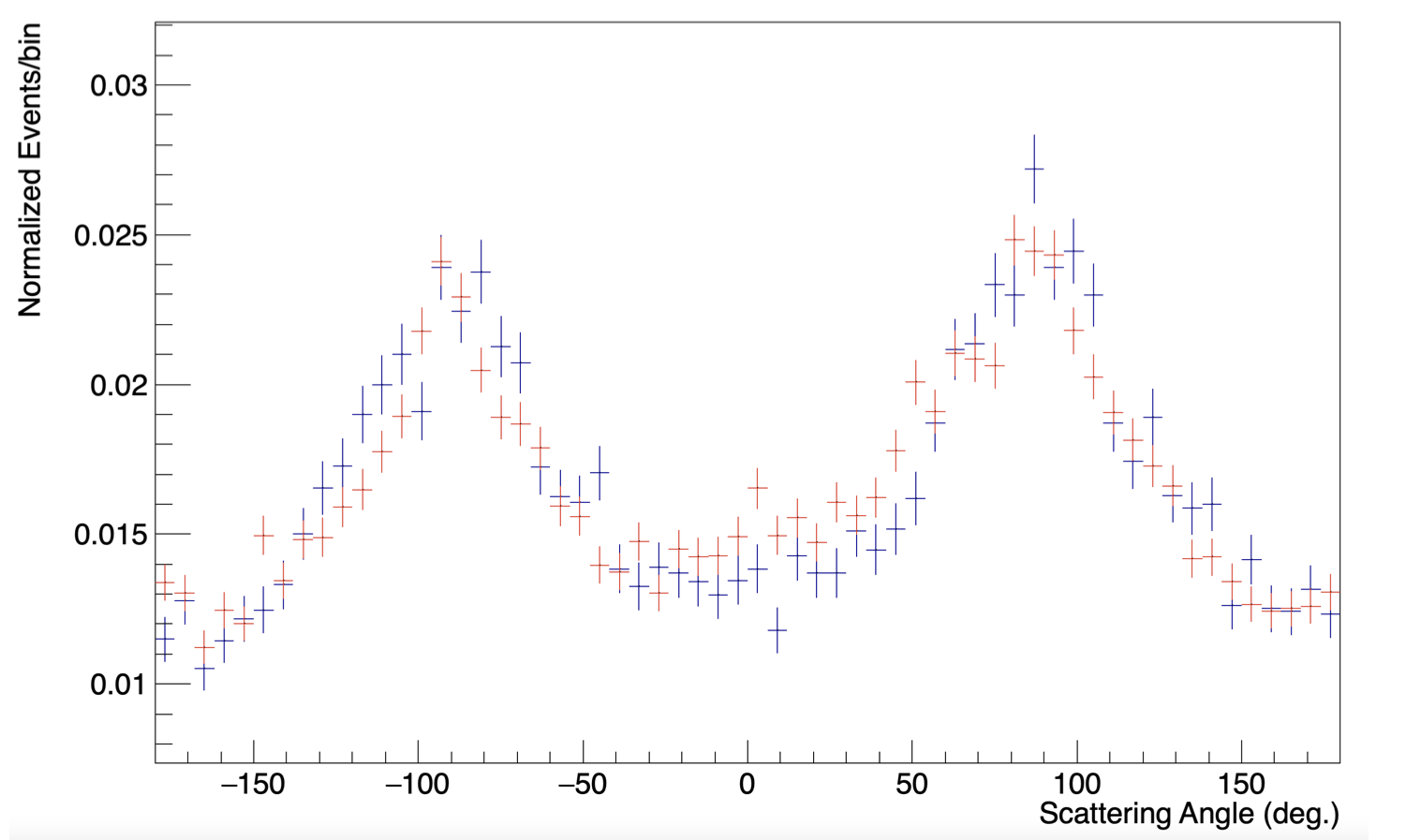}
         \caption{60keV, PD-100\% and PA=0\%}
         \label{fig:60keV_PD100_PA0}
     \end{subfigure}
     \hfill
     \begin{subfigure}[b]{0.48\textwidth}
         \centering
         \includegraphics[scale=0.3]{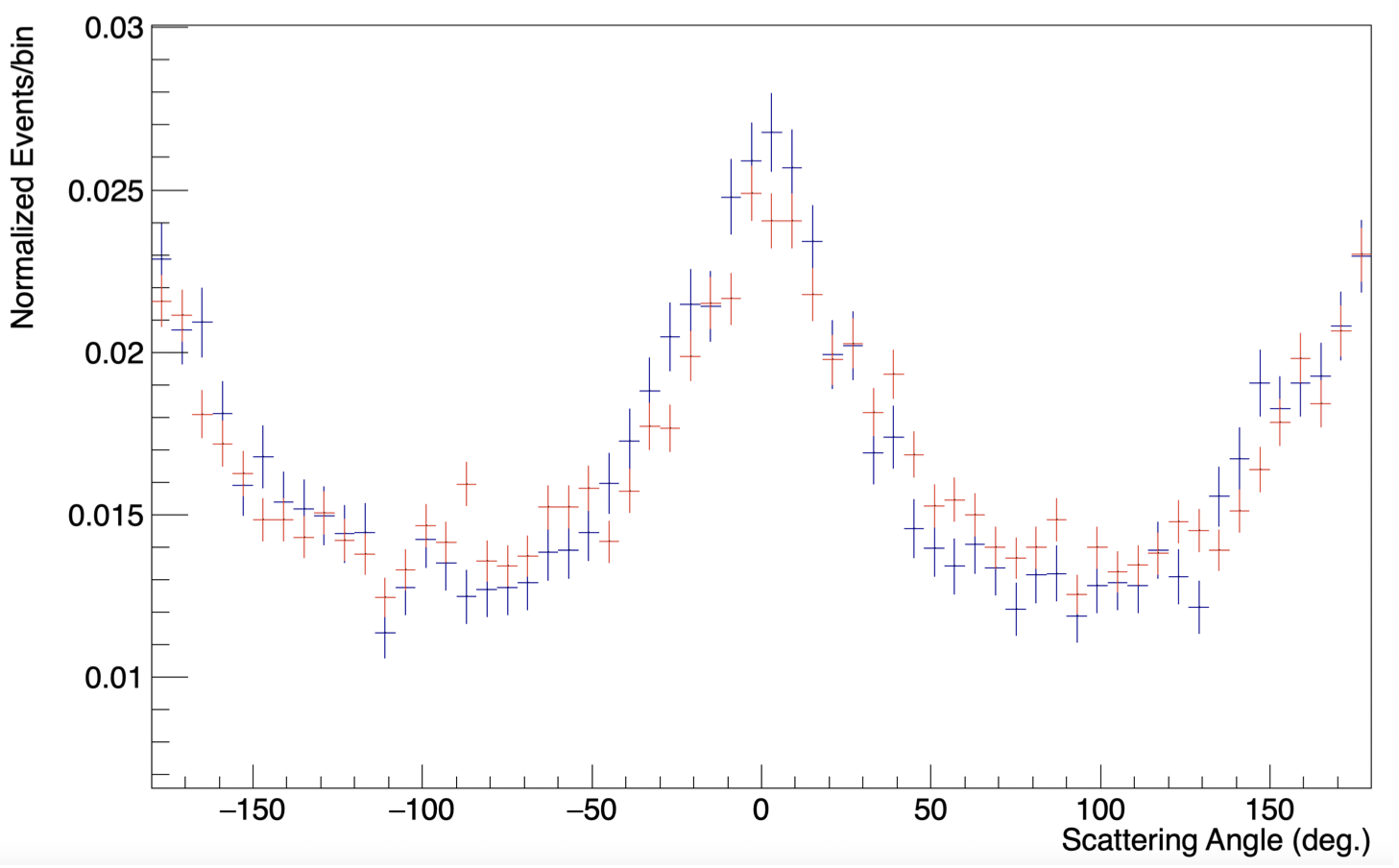}
         \caption{60keV, PD-100\% and PA=90\%}
         \label{fig:60keV_PD100_PA90}
     \end{subfigure}
     \centering
     \begin{subfigure}[b]{0.48\textwidth}
         \centering
         \includegraphics[scale=0.3]{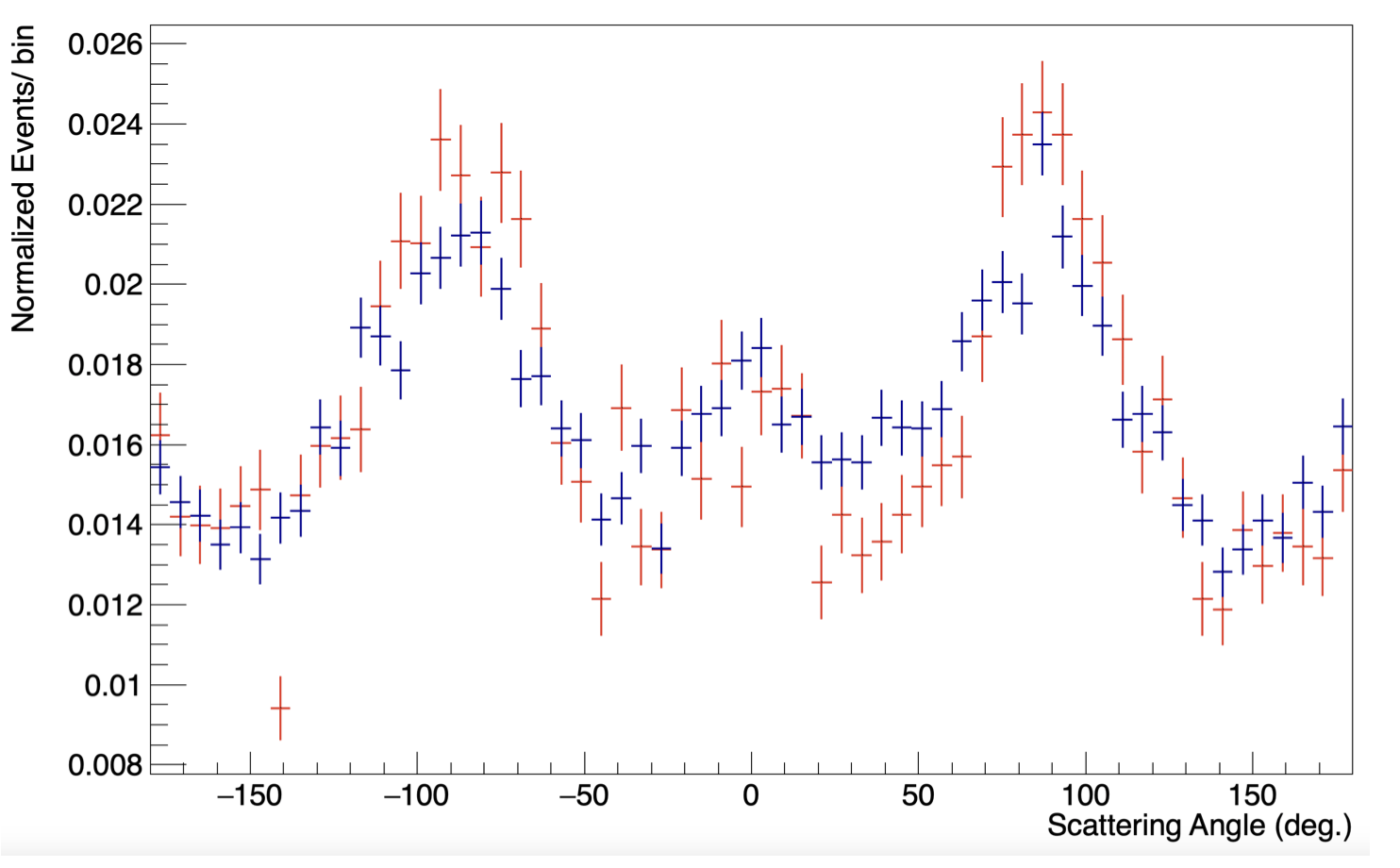}
         \caption{40keV, PD-100\% and PA=0\%}
         \label{fig:40keV_PD100_PA0}
     \end{subfigure}
     \hfill
     \begin{subfigure}[b]{0.48\textwidth}
         \centering
         \includegraphics[scale=0.3]{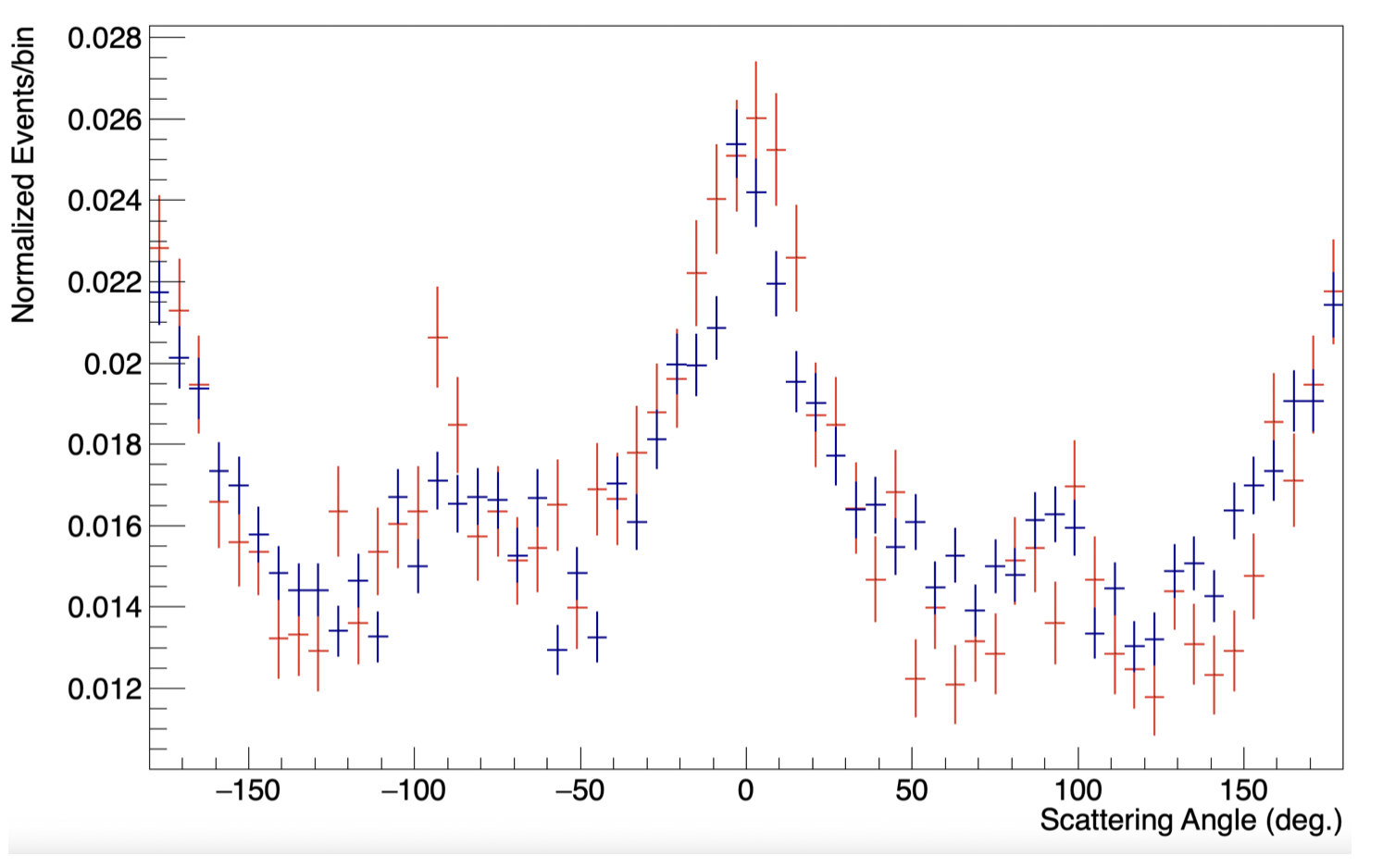}
         \caption{40keV, PD-100\% and PA=90\%}
         \label{fig:40keV_PD100_PA90}
     \end{subfigure}
    \caption{A selection of the results from the ESRF beam test campaign for a single POLAR-2 polarimeter module. The results show a good correspondence of the simulated detector response (blue) with respect to the measured data (red) (figures taken with permissions from \cite{Kole_2024}).}
    \label{fig:ESRFresults}
\end{figure}

The POLAR-2 polarimeter design and simulation package was validated with a single module at the European Synchrotron Radiation Facility (ESRF), which provided 1mm collimated and fully polarized gamma-ray beams with energies of 40 to 120 keV. The polarimeter was placed on a precision x-y translation table to scan on a channel-by-channel basis and study their respective light yield (LY), position-dependent response, and polarization sensitivity. Overall, the results showed that the design goals were achieved and an improvement over POLAR\cite{Kole_2024}. The LY was found to be at approximately 1.5p.e./keV (x5 higher than for POLAR). The LY remained equivalent irrespective of the interaction point in the scintillator bar, validating uniformity across the module. Also, the optical crosstalk was measured to be at 2–3\% for neighboring channels (x10 lower than POLAR). On-axis modulation measurements using 40 and 60 keV beams confirmed the sensitivity of the polarimeter module to gamma-ray polarization and were consistent with the simulated modulation curves (as shown in the Figure \ref{fig:ESRFresults}), indicating the preliminary validity of the simulation framework. 

\begin{figure}
     \centering
     \begin{subfigure}[b]{0.48\textwidth}
         \centering
         \includegraphics[scale=0.5]{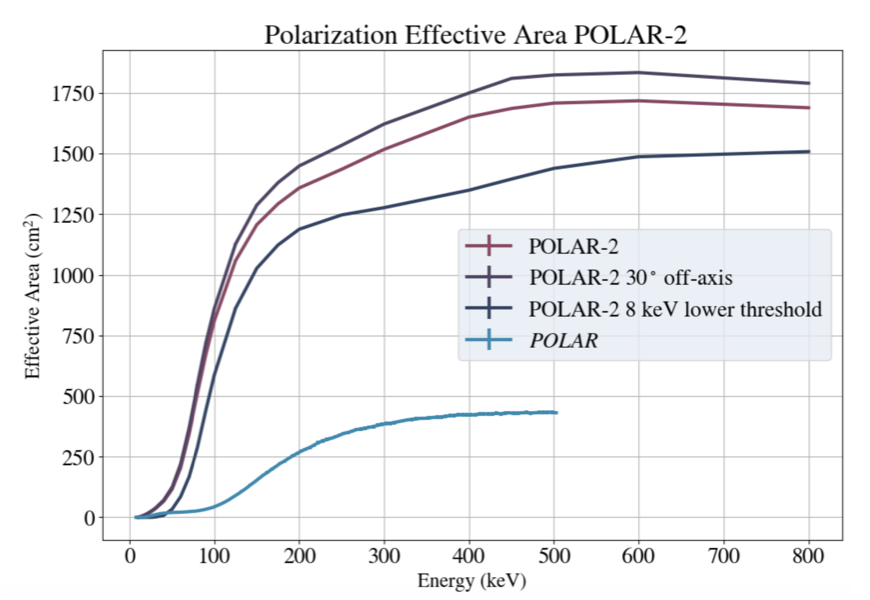}
         \caption{}
         \label{fig:polar2_arf}
     \end{subfigure}
     \hfill
     \begin{subfigure}[b]{0.48\textwidth}
         \centering
         \includegraphics[scale=0.32]{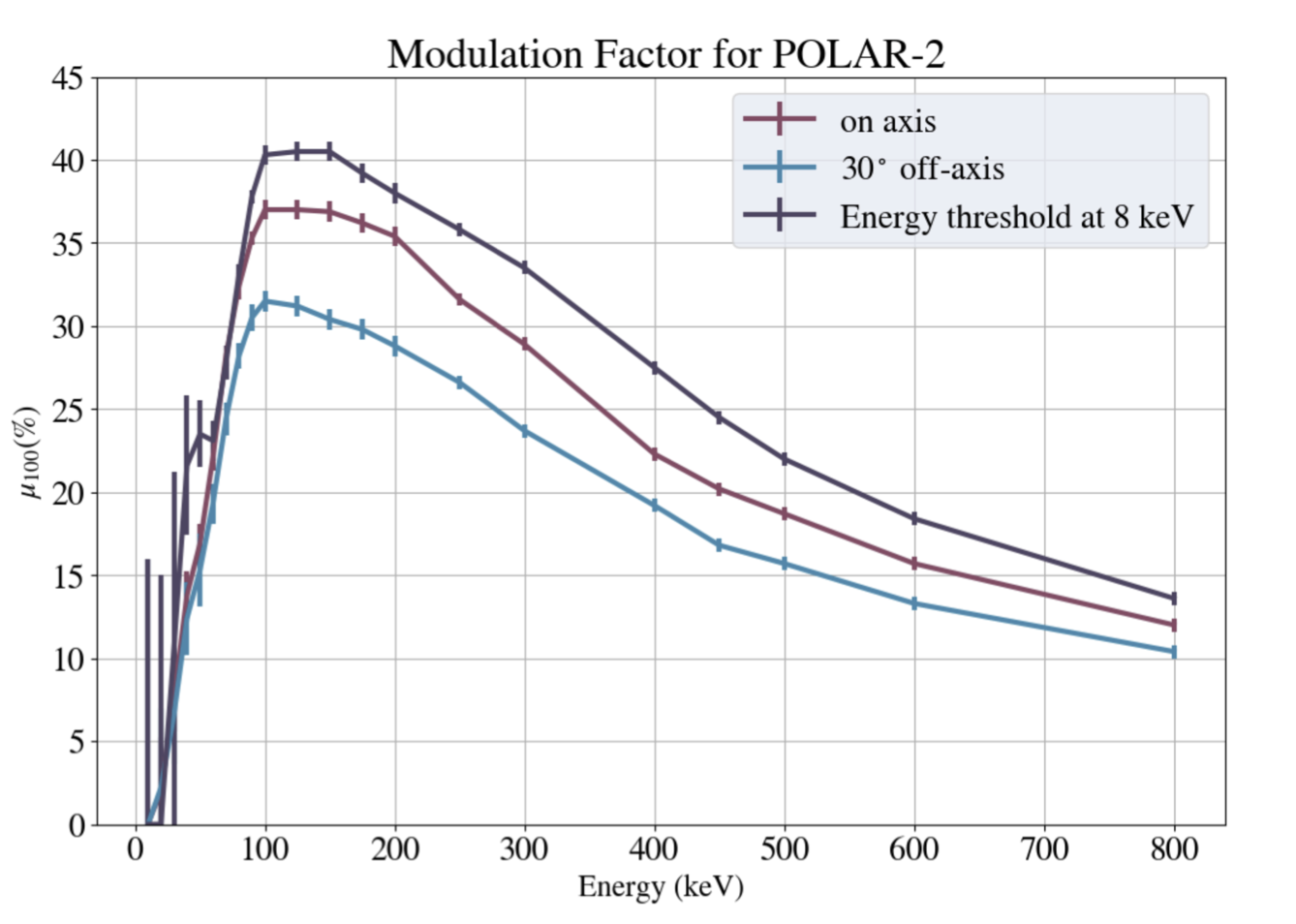}
         \caption{}
         \label{fig:polar2_modulationfactor}
     \end{subfigure}
    \caption{Expected POLAR-2 \textbf{a)} effective area for polarized events and \textbf{b)} modulation factor inferred from the ESRF beam test (figures taken with permissions from \cite{Kole_2024})}
    \label{fig:polar2_arf_m100}
\end{figure}

Based on the ESRF results, the effective area of POLAR-2 is projected to exceed 100 $cm^2$ at 40 keV, where POLAR only reached this sensitivity above 120 keV. This enhanced effective area is critical for detecting a broader range of gamma-ray bursts (GRBs), particularly those with low-energy emissions. The modulation factor ($\mu_{100}$), a critical parameter for polarization measurements, was also analyzed for the full instrument (see the figures in \Cref{fig:polar2_arf_m100}). With a higher modulation factor and reduced systematic uncertainties compared to POLAR, POLAR-2 is expected to deliver unprecedented precision in measuring the degree and angle of polarization in GRB prompt emissions. Below 200 keV the modulation factor is about 10\% points higher for POLAR-2 compared to POLAR. The achievements imply that POLAR-2 will be able to accomplish its intended scientific goals, especially for time-resolved and energy-resolved polarization studies (as outlined at the beginning of this chapter) \cite{polar2design_prospects}.

\section{Calibration Module}

\begin{figure}[h!]
     \centering
     \begin{subfigure}[b]{0.32\textwidth}
         \centering
         \includegraphics[scale=0.035]{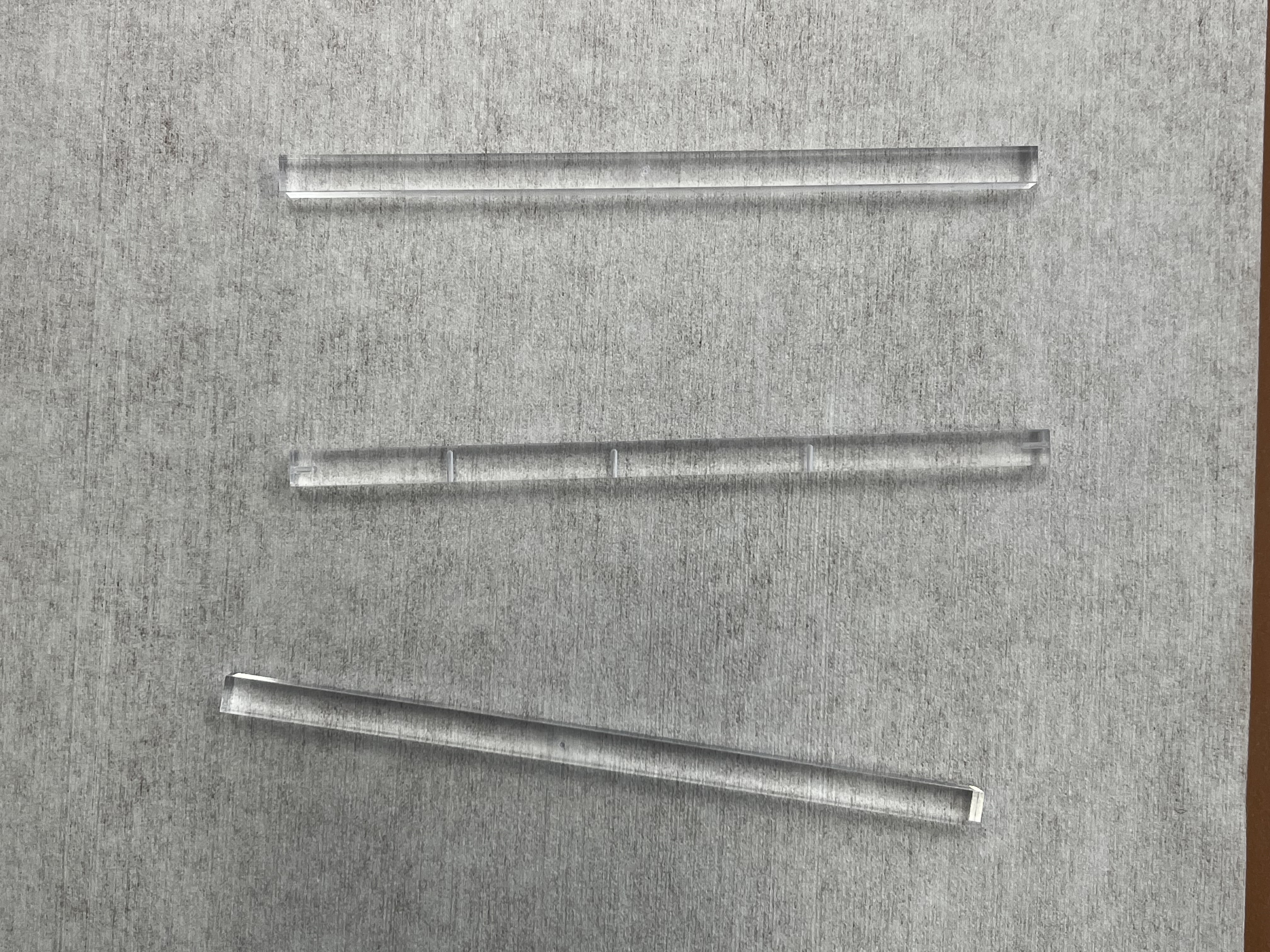}
         \caption{}
         \label{fig:bars_drilled}
     \end{subfigure}
     \hfill
     \begin{subfigure}[b]{0.32\textwidth}
         \centering
         \includegraphics[scale=0.035]{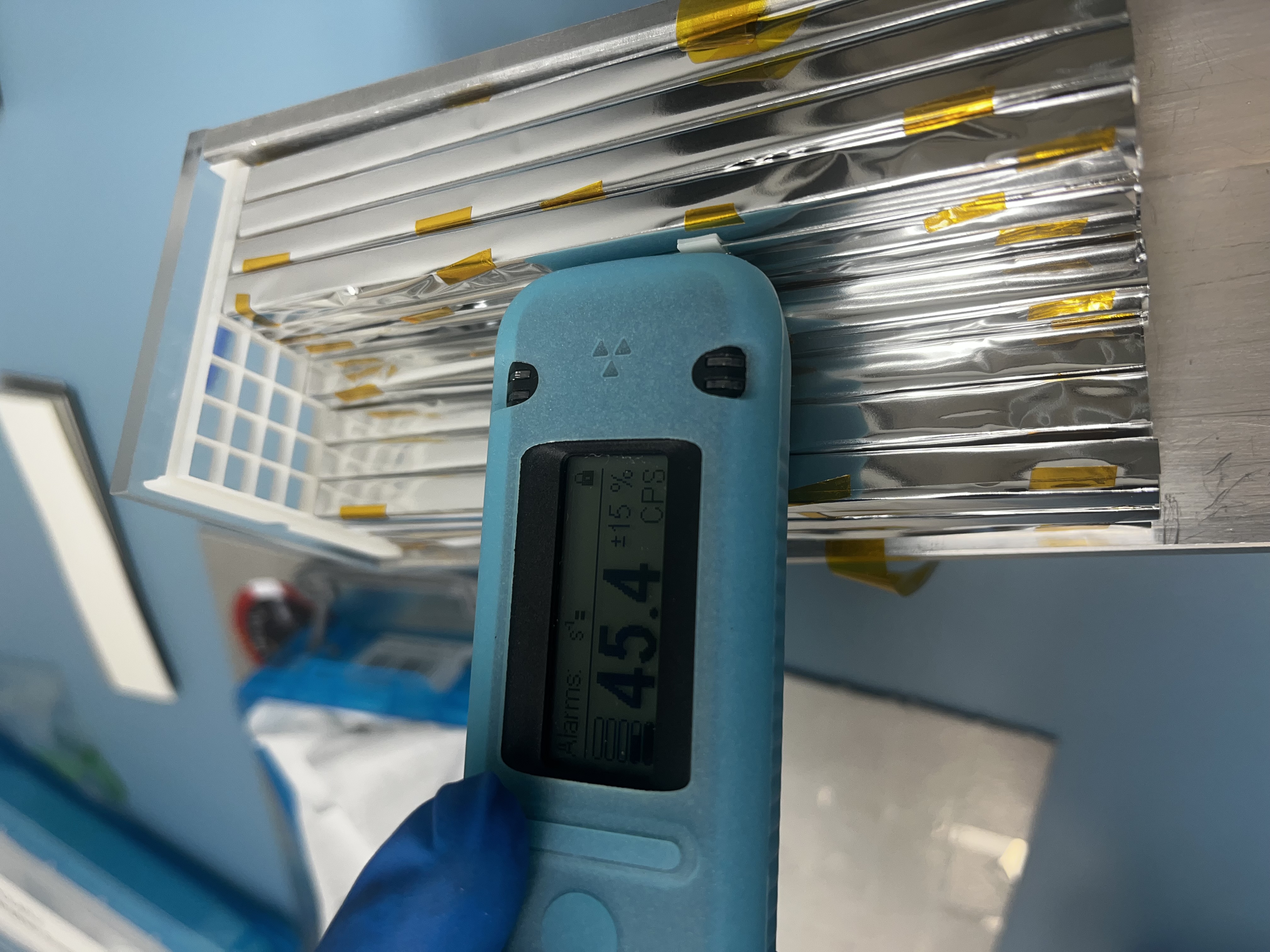}
         \caption{}
         \label{fig:tagged_bar}
     \end{subfigure}
     \centering
     \begin{subfigure}[b]{0.32\textwidth}
         \centering
         \includegraphics[scale=0.26]{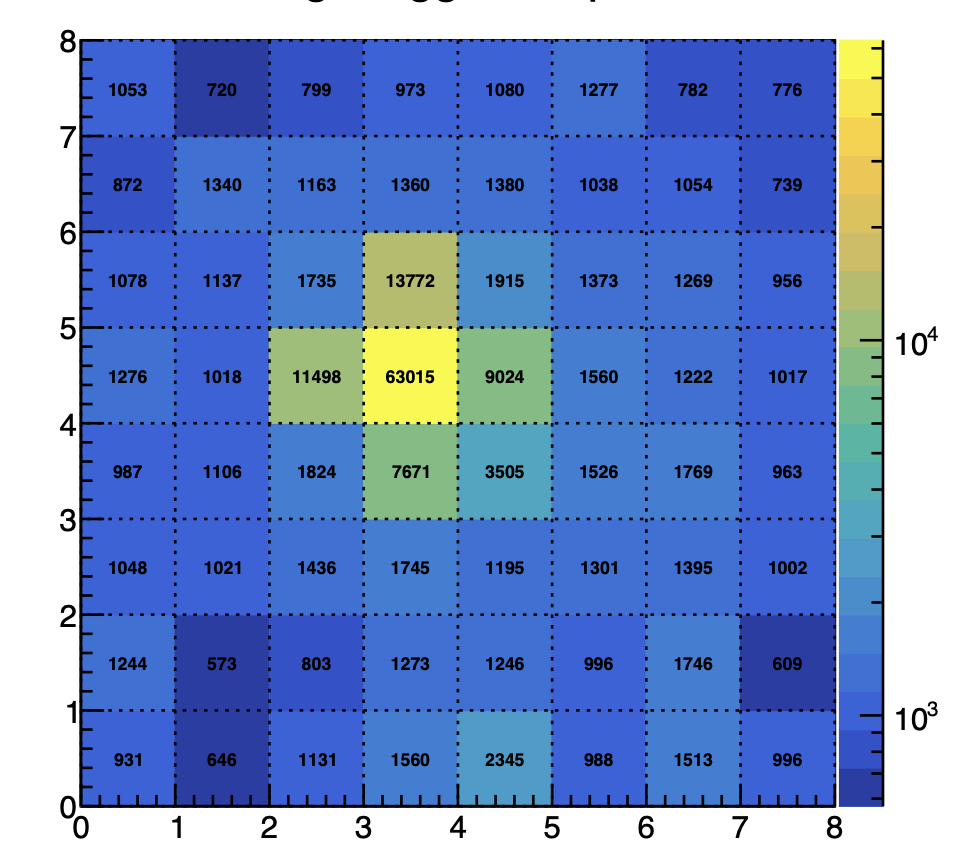}
         \caption{}
         \label{fig:hitmap_taggedmod}
     \end{subfigure}
    \caption{\textbf{a)} Photo showing the size of the hole when drilled inside the scintillator compared to a standard scintillator bar. On the center hole is drilled in the tagged module. \textbf{b)} Activity of the "tagged" bar measured with a gamma detector after it has been placed wrapped and assembled inside module. \textbf{c)} Hit map of the Am-241 inside the "tagged" module, highlighting "tagged" bar.}
    \label{fig:taggedmodule}
\end{figure}

As POLAR-2 is expected to operate for at least two years, an in-orbit calibration strategy is essential. For that purpose, a \emph{calibration module} has been developed, incorporating an Am-241 radioactive source embedded within one of the scintillator bars. Here, a small Am-241 source of about $\sim$200Bq is placed inside a \SI{1}{mm} hole in the scintillator (see example in the photo of \ref{fig:bars_drilled}). The hole is then sealed with a plastic scintillator cap of the same material, therefore maximizing the light produced by the decay products. The Am-241 will emit an alpha particle inside its bar, whereas the \SI{59.5}{keV} gamma will mostly propagate through the scintillators, yielding a distinct signature which can be used for calibration. Not all modules will be associated with such a "tagged" bar. The exact number, final chosen activity and distribution inside the full polarimeter is still under investigation. A prototype "tagged" module is currently undergoing extensive testing to study gain stability and temperature dependence, aiming to mitigate issues caused by temperature fluctuations during space operations. The hit map of the module, shown in Figure \ref{fig:hitmap_taggedmod}, depicts the highest count rate in the scintillator bar containing the embedded source, with elevated counts in neighboring bars due to optical crosstalk.

\section{Qualification and Production}

The POLAR-2 prototype modules were subjected to extensive qualification campaigns, including radiation campaigns \cite{Mianowski2023,Mianowski2023b}, vibration and shock tests, and thermal cycling. A mitigation strategy—low temperature operation combined with periodic annealing \cite{DEANGELIS2023167934}—has been devised to suppress radiation induced noise, allowing it operate beyond the nominal two-year mission lifetime. The procurement of long lead items is nearly completed. Most of the scintillator bars have been delivered and all the SiPM arrays have been delivered and characterized. For every scintillator batch, a sample of 20 units is examined: the dimension as well as surface quality is checked to be sure they comply with the requirements. The CFRP sockets are being procured. Also the production of a qualification model composed of \numproduct{3x3} will be completed  before the end of 2025 and tested at a follow-up ESRF beam campaign in 2026. The 9 tested modules will then be integrated into a larger \numproduct{10x10} test model, together with 91 \emph{mechanical} modules (with 'heating' PCBs) which will be used for vibration and thermal vacuum tests in 2026. Mission preparations are proceeding at full pace, with the interface definition between POLAR-2 and the spacecraft control system (CSS) approaching finalization. 

\acknowledgments
We gratefully acknowledge the Swiss Space Office of the State Secretariat for Education, Research and Innovation (ESA PRODEX Programme) which supported the development and production of the POLAR-2 detector. National Centre for Nuclear Research acknowledges support from Polish National Science Center under the grant UMO-2018/30/M/ST9/00757. We gratefully acknowledge the support from the National Natural Science Foundation of China (Grant No. 11961141013, 11503028), the Xie Jialin Foundation of the Institute of High Energy Physics, Chinese Academy of Sciences (Grant No. 2019IHEPZZBS111), the Joint Research Fund in Astronomy under the cooperative agreement between the National Natural Science Foundation of China and the Chinese Academy of Sciences (Grant No. U1631242), the National Basic Research Program (973 Program) of China (Grant No. 2014CB845800), the Strategic Priority Research Program of the Chinese Academy of Sciences (Grant No. XDB23040400), and the Youth Innovation Promotion Association of Chinese Academy of Sciences.

\bibliographystyle{JHEP}
\small
\bibliography{bibliography20250928}

@article{PRODUIT2018259,
title = {{Design and construction of the POLAR detector}},
journal = {Nuclear Instruments and Methods in Physics Research Section A: Accelerators, Spectrometers, Detectors and Associated Equipment},
volume = {877},
pages = {259-268},
year = {2018},
issn = {0168-9002},
doi = {https://doi.org/10.1016/j.nima.2017.09.053},
url = {https://www.sciencedirect.com/science/article/pii/S0168900217310239},
author = {N. Produit and T.W. Bao and T. Batsch and T. Bernasconi and I. Britvich and F. Cadoux and I. Cernuda and J.Y. Chai and Y.W. Dong and N. Gauvin and W. Hajdas and M. Kole and M.N. Kong and R. Kramert and L. Li and J.T. Liu and X. Liu and R. Marcinkowski and S. Orsi and M. Pohl and D. Rapin and D. Rybka and A. Rutczynska and H.L. Shi and P. Socha and J.C. Sun and L.M. Song and J. Szabelski and I. Traseira and H.L. Xiao and R.J. Wang and X. Wen and B.B. Wu and L. Zhang and L.Y. Zhang and S.N. Zhang and Y.J. Zhang and A. Zwolinska}
}

@article{lundman2016polarizationgammarayburstsdissipative,
      title={Polarization of gamma-ray bursts in the dissipative photosphere model}, 
      author={Christoffer Lundman and Indrek Vurm and Andrei M. Beloborodov},
      year={2016},
      eprint={1611.01451},
      archivePrefix={arXiv},
      volume = {1611.01451},
      journal={arXiv},
      primaryClass={astro-ph.HE},
      doi={10.48550/arXiv.1611.01451},
      url={https://arxiv.org/abs/1611.01451}, 
}

@Article{galaxies9040082,
AUTHOR = {Gill, Ramandeep and Kole, Merlin and Granot, Jonathan},
TITLE = {{GRB Polarization: A Unique Probe of GRB Physics}},
JOURNAL = {Galaxies},
VOLUME = {9},
YEAR = {2021},
NUMBER = {4},
ARTICLE-NUMBER = {82},
URL = {https://www.mdpi.com/2075-4434/9/4/82},
ISSN = {2075-4434},
DOI = {10.3390/galaxies9040082}
}

@article{Produit_2018,
   title={{Design and construction of the POLAR detector}},
   volume={877},
   ISSN={0168-9002},
   url={http://dx.doi.org/10.1016/j.nima.2017.09.053},
   DOI={10.1016/j.nima.2017.09.053},
   journal={Nuclear Instruments and Methods in Physics Research Section A: Accelerators, Spectrometers, Detectors and Associated Equipment},
   publisher={Elsevier BV},
   author={Produit, N. and Bao, T.W. and Batsch, T. and Bernasconi, T. and Britvich, I. and Cadoux, F. and Cernuda, I. and Chai, J.Y. and Dong, Y.W. and Gauvin, N. and Hajdas, W. and Kole, M. and Kong, M.N. and Kramert, R. and Li, L. and Liu, J.T. and Liu, X. and Marcinkowski, R. and Orsi, S. and Pohl, M. and Rapin, D. and Rybka, D. and Rutczynska, A. and Shi, H.L. and Socha, P. and Sun, J.C. and Song, L.M. and Szabelski, J. and Traseira, I. and Xiao, H.L. and Wang, R.J. and Wen, X. and Wu, B.B. and Zhang, L. and Zhang, L.Y. and Zhang, S.N. and Zhang, Y.J. and Zwolinska, A.},
   year={2018},
   month=jan, pages={259–268} }

@article{Kole2020,
   author = {M. Kole and N. De Angelis and F. Berlato and J. M. Burgess and N. Gauvin and J. Greiner and W. Hajdas and H. C. Li and Z. H. Li and A. Pollo and N. Produit and D. Rybka and L. M. Song and J. C. Sun and J. Szabelski and T. Tymieniecka and Y. H. Wang and B. B. Wu and X. Wu and S. L. Xiong and S. N. Zhang and Y. J. Zhang},
   doi = {10.1051/0004-6361/202037915},
   issn = {14320746},
   journal = {Astronomy and Astrophysics},
   month = {12},
   publisher = {EDP Sciences},
   title = {{The POLAR gamma-ray burst polarization catalog}},
   volume = {644},
   year = {2020}
}

@Article{Zhang2019,
author={Zhang, Shuang-Nan
and Kole, Merlin
and Bao, Tian-Wei
and Batsch, Tadeusz
and Bernasconi, Tancredi
and Cadoux, Franck
and Chai, Jun-Ying
and Dai, Zi-Gao
and Dong, Yong-Wei
and Gauvin, Neal
and Hajdas, Wojtek
and Lan, Mi-Xiang
and Li, Han-Cheng
and Li, Lu
and Li, Zheng-Heng
and Liu, Jiang-Tao
and Liu, Xin
and Marcinkowski, Radoslaw
and Produit, Nicolas
and Orsi, Silvio
and Pohl, Martin
and Rybka, Dominik
and Shi, Hao-Li
and Song, Li-Ming
and Sun, Jian-Chao
and Szabelski, Jacek
and Tymieniecka, Teresa
and Wang, Rui-Jie
and Wang, Yuan-Hao
and Wen, Xing
and Wu, Bo-Bing
and Wu, Xin
and Wu, Xue-Feng
and Xiao, Hua-Lin
and Xiong, Shao-Lin
and Zhang, Lai-Yu
and Zhang, Li
and Zhang, Xiao-Feng
and Zhang, Yong-Jie
and Zwolinska, Anna},
title={Detailed polarization measurements of the prompt emission of five gamma-ray bursts},
journal={Nature Astronomy},
year={2019},
month={Mar},
day={01},
volume={3},
number={3},
pages={258-264},
issn={2397-3366},
doi={10.1038/s41550-018-0664-0},
url={https://doi.org/10.1038/s41550-018-0664-0}
}

@inproceedings{DeAngelis:2023E/,
  author = "De Angelis, Nicolas  and  Burgess, J. Michael  and  Cadoux, Franck  and  Greiner, Jochen  and  Kole, Merlin  and  Li, Hancheng  and  Mianowski, Slawomir  and  Pollo, Agnieszka  and  Produit, Nicolas  and  Rybka, Dominik  and  Sun, Jianchao  and  Wu, Xin  and  Zhang, Shuang-Nan",
  title = "{Energy-dependent polarization of Gamma-Ray Bursts' prompt emission with the POLAR and POLAR-2 instruments}",
  doi = "10.22323/1.444.0619",
  booktitle = "Proceedings of 38th International Cosmic Ray Conference {\textemdash} PoS(ICRC2023)",
  year = 2023,
  volume = "444",
  pages = "619"
}

@Article{Mianowski2023,
author={Mianowski, Slawomir
and De Angelis, Nicolas
and Hulsman, Johannes
and Kole, Merlin
and Kowalski, Tomasz
and Kusyk, Sebastian
and Li, Hancheng
and Mianowska, Zuzanna
and Mietelski, Jerzy
and Pollo, Agnieszka
and Rybka, Dominik
and Sun, Jianchao
and Swakon, Jan
and Wrobel, Damian
and Wu, Xin},
title={{Proton irradiation of SiPM arrays for POLAR-2}},
journal={Experimental Astronomy},
year={2023},
month={Apr},
day={01},
volume={55},
number={2},
pages={343-371},
issn={1572-9508},
doi={10.1007/s10686-022-09873-6},
url={https://doi.org/10.1007/s10686-022-09873-6}
}

@Article{Mianowski2023b,
author={Mianowski, Slawomir
and De Angelis, Nicolas
and Brylew, Kamil
and Hulsman, Johannes
and Kowalski, Tomasz
and Kusyk, Sebastian
and Mianowska, Zuzanna
and Mietelski, Jerzy
and Rybka, Dominik
and Swakon, Jan
and Wrobel, Damian},
title={{Proton irradiation of plastic scintillator bars for POLAR-2}},
journal={Experimental Astronomy},
year={2023},
month={Dec},
day={01},
volume={56},
number={2},
pages={355-370},
issn={1572-9508},
doi={10.1007/s10686-023-09906-8},
url={https://doi.org/10.1007/s10686-023-09906-8}
}

@article{DEANGELIS2023167934,
title = {{Temperature dependence of radiation damage annealing of Silicon Photomultipliers}},
journal = {Nuclear Instruments and Methods in Physics Research Section A: Accelerators, Spectrometers, Detectors and Associated Equipment},
volume = {1048},
pages = {167934},
year = {2023},
issn = {0168-9002},
doi = {https://doi.org/10.1016/j.nima.2022.167934},
url = {https://www.sciencedirect.com/science/article/pii/S0168900222012268},
author = {N. {De Angelis} and M. Kole and F. Cadoux and J. Hulsman and T. Kowalski and S. Kusyk and S. Mianowski and D. Rybka and J. Stauffer and J. Swakon and D. Wrobel and X. Wu}
}

@article{Kole_2024,
   title={{Response of the first POLAR-2 prototype to polarized beams}},
   volume={19},
   ISSN={1748-0221},
   url={http://dx.doi.org/10.1088/1748-0221/19/08/P08002},
   DOI={10.1088/1748-0221/19/08/p08002},
   number={08},
   journal={Journal of Instrumentation},
   publisher={IOP Publishing},
   author={Kole, Merlin and Angelis, Nicolas De and Bacelj, Ana and Cadoux, Franck and Elwertowska, Agnieszka and Hulsman, Johannes and Li, Hancheng and Łubian, Grzegorz and Kowalski, Tomasz and Koziol, Gilles and Pollo, Agnieszka and Produit, Nicolas and Rybka, Dominik and Stil, Adrien and Sun, Jianchao and Wu, Xin and Zezuliński, Kacper and Zhang, Shuang-Nan},
   year={2024},
   month=aug, pages={P08002} }

@article{Kole_2025,
   title={{Design and performance of a universal SiPM readout system for X- and gamma-ray missions}},
   volume={1080},
   ISSN={0168-9002},
   url={http://dx.doi.org/10.1016/j.nima.2025.170782},
   DOI={10.1016/j.nima.2025.170782},
   journal={Nuclear Instruments and Methods in Physics Research Section A: Accelerators, Spectrometers, Detectors and Associated Equipment},
   publisher={Elsevier BV},
   author={Kole, Merlin and De Angelis, Nicolas and Produit, Nicolas and Cadoux, Franck and Favre, Yannick and Greiner, Jochen and Hulsman, Johannes and Kusyk, Sebastian and Li, Hancheng and Rybka, Dominik and Stauffer, Jerome and Stil, Adrien and Sun, Jianchao and Swakon, Jan and Wrobel, Damian and Wu, Xin},
   year={2025},
   month=nov, pages={170782} }

@article{DeAngelis_2025,
doi = {10.1088/1748-0221/20/02/P02010},
url = {https://doi.org/10.1088/1748-0221/20/02/P02010},
year = {2025},
month = {feb},
publisher = {IOP Publishing},
volume = {20},
number = {02},
pages = {P02010},
author = {De Angelis, Nicolas and Cadoux, Franck and Husi, Coralie and Kole, Merlin and Mianowski, Sławomir},
title = {{Optimizing the light output of a plastic scintillator and SiPM based detector through optical characterization and simulation: a case study for POLAR-2}},
journal = {Journal of Instrumentation}
}

@inproceedings{Produit2023,
  series = {ICRC2023},
  title = {POLAR-2,  the next generation of GRB polarization detector},
  url = {http://dx.doi.org/10.22323/1.444.0550},
  DOI = {10.22323/1.444.0550},
  booktitle = {Proceedings of 38th International Cosmic Ray Conference — PoS(ICRC2023)},
  publisher = {Sissa Medialab},
  author = {Produit,  Nicolas and Kole,  Merlin and Wu,  Xin and De Angelis,  Nicolas and Li,  Hancheng and Rybka,  Dominik and Pollo,  Agnieszka and Mianowski,  Slawomir and Greiner,  Jochen and Burgess,  J Michael and Sun,  Jianchao and Zhang,  Shuang-Nan},
  year = {2023},
  month = aug,
  pages = {550},
  collection = {ICRC2023}
}

@misc{polar2design_prospects,
  doi = {10.48550/ARXIV.2510.02016},
  url = {https://arxiv.org/abs/2510.02016},
  author = {Kole,  Merlin and de Angelis,  Nicolas and He,  Jiang and Liu,  Hongbang and Sun,  Jianchao and Xie,  Fei and Zaid,  Jimmy},
  title = {Design and Scientific Prospects of the POLAR-2 Mission},
  publisher = {arXiv},
  year = {2025},
  copyright = {Creative Commons Attribution 4.0 International}
}

@article{Burgess2019,
  title = {Time-resolved GRB polarization with POLAR and GBM: Simultaneous spectral and polarization analysis with synchrotron emission},
  volume = {627},
  ISSN = {1432-0746},
  url = {http://dx.doi.org/10.1051/0004-6361/201935056},
  DOI = {10.1051/0004-6361/201935056},
  journal = {Astronomy and Astrophysics},
  publisher = {EDP Sciences},
  author = {Burgess,  J. M. and Kole,  M. and Berlato,  F. and Greiner,  J. and Vianello,  G. and Produit,  N. and Li,  Z. H. and Sun,  J. C.},
  year = {2019},
  month = jul,
  pages = {A105}
}

@article{Yonetoku2011,
  title = {Gamma-Ray Burst Polarimeter (GAP) aboard the Small Solar Power Sail Demonstrator IKAROS},
  volume = {63},
  ISSN = {0004-6264},
  url = {http://dx.doi.org/10.1093/pasj/63.3.625},
  DOI = {10.1093/pasj/63.3.625},
  number = {3},
  journal = {Publications of the Astronomical Society of Japan},
  publisher = {Oxford University Press (OUP)},
  author = {Yonetoku,  Daisuke and Murakami,  Toshio and Gunji,  Shuichi and Mihara,  Tatehiro and Sakashita,  Tomonori and Morihara,  Yoshiyuki and Kikuchi,  Yukihiro and Takahashi,  Takuya and Fujimoto,  Hirofumi and Toukairin,  Noriyuki and Kodama,  Yoshiki and Kubo,  Shin},
  year = {2011},
  month = jun,
  pages = {625–638}
}

@misc{de_angelis_2023_thesis,
  author       = {De Angelis, Nicolas},
  title        = {Development of the Next Generation Space-based Compton Polarimeter and Energy Resolved Polarization Analysis of Gamma-Ray Bursts Prompt Emission},
  month        = dec,
  year         = 2023,
}

\end{document}